\documentclass[journal,peerreview]{IEEEtran}
\usepackage{cite}
\ifCLASSINFOpdf
  \usepackage[pdftex]{graphicx}
\else
  \usepackage[dvips]{graphicx}
  \graphicspath{{../figures/}}
\fi
\usepackage[dvipsnames]{xcolor}

\usepackage{amsmath}
\usepackage{mathrsfs}
\usepackage{mathtools}
\usepackage{array}
\ifCLASSOPTIONcompsoc
  \usepackage[caption=false,font=normalsize,labelfont=sf,textfont=sf]{subfig}
\else
  \usepackage[caption=false,font=footnotesize]{subfig}
\fi

\usepackage{soul}
\usepackage{xcolor}

\newcounter{MYtempeqncnt}

\begin{document}
\title{Coherent Multi-Transducer Ultrasound Imaging}

\author{Laura~Peralta,
        Alberto~Gomez,
         Ying~Luan, Baehyung~Kim, Joseph~V~Hajnal
        and~Robert~J~Eckersley % <-this % stops a space
\thanks{L. Peralta, A. Gomez, B. Kim, Y. Luan, J. Hajnal and R. Eckersley are with the Department of Biomedical Engineering, School of Biomedical Engineering \& Imaging Sciences,
King's College London, King's Health Partners, St Thomas' Hospital, London, SE1 7EH, United
Kingdom (e-mail: laura.peralta-pereira@kcl.ac.uk).}% <-this % stops a space
\thanks{Y. Luan is with GE Healthcare.}% <-this % stops a space
\thanks{B. Kim is with Ultrasound Research Laboratory, Physiology and Biomedical Engineering, Mayo Clinic College of Medicine, Rochester, USA.}% <-this % stops a space
}


\markboth{IEEE TRANSACTIONS ON ULTRASONICS, FERROELECTRICS, AND FREQUENCY CONTROL, July~2018}%
{Peralta \MakeLowercase{\textit{et al.}}: Coherent Multi-transducer Ultrasound Imaging}



\maketitle


\begin{abstract}
An extended aperture has the potential to greatly improve ultrasound imaging performance. This work extends the effective aperture size by coherently compounding the received radio frequency data from multiple transducers. A framework is developed in which an ultrasound imaging system consisting of $N$ synchronized matrix arrays, each with partly shared field of view, take turns to transmit plane waves. Only one individual transducer transmits at each time while all $N$ transducers simultaneously receive.   
The subwavelength localization accuracy required to combine information from multiple transducers is achieved without the use of any external tracking device. 
The method developed in this study is based on the study of the backscattered echoes received by the same transducer and resulting from a targeted scatterer point in the medium insonated by the multiple ultrasound probes of the system. The current transducer locations along with the speed of sound in the medium are deduced by optimizing the cross-correlation between these echoes. 
The method is demonstrated experimentally in 2-D using ultrasound point and anechoic lesion phantoms and a first demonstration of a free-hand experiment is also shown. Results demonstrate that the coherent multi-transducer imaging has the potential to improve ultrasound image quality, improving resolution and target detectability. Lateral resolution, contrast and contrast-to-noise ratio improved from 0.67 mm, -6.708 dB and 0.702, respectively, when using a single probe, to 0.18 mm, -7.251 dB and 0.721 in the coherent multi-transducer imaging case.

\end{abstract}

\begin{IEEEkeywords}
Mulit-probe, Image resolution, Large aperture, Plane Wave, Ultrasound imaging  
\end{IEEEkeywords}

\IEEEpeerreviewmaketitle

%\ifCLASSOPTIONpeerreview
%\begin{center} \bfseries EDICS Category: 3-BBND \end{center}
%\fi

\section{Introduction}


\IEEEPARstart{U}{ltrasound} is a widely used clinical imaging tool and its advantages in terms of safety and low cost over other medical imaging modalities are well known.  However, conventional ultrasound systems yield images which can be difficult to assess, because of the limited resolution and view-dependent artefacts that are inherent to the small aperture transducers used clinically, particularly at larger depths in abdominal or fetal imaging applications. To increase the field of view (FoV), multiple images can be incoherently compounding together in the lateral direction using image registration or mechanically moving the probe \cite{weng1997us, zimmer2018}. Notwithstanding, the resolution of the resulting image is not improved by such approaches. However, an extended aperture has the potential to greatly improve resolution and imaging performance \cite{moshfeghi1988vivo}. 
   
Recently, the improvements of a wider coherent aperture have been shown in synthetic aperture ultrasound imaging \cite{bottenus2017evaluation, zhang2016synthetic}, where an extended aperture was obtained by mechanically moving and tracking the ultrasound transducer. The tracking information was used to identify the relative position and orientation of the ultrasound images which were then merged together into a final image. 
However, noise in the tracking system and calibration errors are propagated to coherent image reconstruction causing image degradation. The subwavelength localization accuracy required to merge information from multiple transducers is challenging to achieve in conventional ultrasound calibration. Resolution will suffer from motion artefacts, tissue deformation or tissue aberration, which worsen with increased effective aperture size \cite{gammelmark20142,zhang2016co}. For practical implementation, a more accurate calibration technique is required \cite{bottenus2016feasibility, zhang2016synthetic}. In addition, the viability of the technique \emph{in-vivo} is limited by the long acquisition times ($>$15 minutes per image) that may also contribute to the break down of the coherent aperture \cite{bottenus2017evaluation}. 

On the other hand, in clinical practice the aperture size is limited not only by the complexity of the system and its high cost but also by the low flexibility that a large probe may have for different situations. Clinical probes must be controlled and moved by a physician to adapt to contours and shapes of the human body. The physical transducer size is then a compromise between cost, ergonomics and image performance. Improving the ultrasound image quality without modifying the shape of conventional ultrasound probes may be of great interest. To demonstrate this concept, this paper describes the coherent combination of conventional transducers, in order to provide a significantly extended aperture. The motivation of this work is to demonstrate the potential of this approach and provide a proof-of-concept as an initial step towards large array imaging using non-continuous extended apertures. The novelty of this work lies in the use of the mutual information available in the signals received by the individual transducers that form the extended array to provide precise relative positioning information so that they might be used as one coherent whole. It is the first time, to our knowledge, that such an approach has been described and demonstrated.

%This work proposes a fully coherent multi-transducer ultrasound imaging system, formed by multiple ultrasound transducers that are synchronized and freely \hl{placed} in space. Through coherent integration of the signals received by  different transducers, a larger effective aperture is obtained, leading to an improved final image. 
Generation of a coherent aperture requires the position of transmitters and receivers to be known to subwavelength accuracy \cite{denarie2013coherent}. The method presented here achieves such an accurate localization. A unique aspect of this approach is that it does not require an external tracking system to achieve accurate localization. Instead, the coherence in the backscattered echoes resulting from point-like scatterers in the medium is used to determine the relative position of multiple transducers with respect to a single transducer.
%The method presented here achieves such an accurate localization, based on the coherence of the backscattered echoes received by the same transducer and resulting from a targeted scatterer point in the medium isonated by the multiple ultrasound probes of the system. This enables the multiple transducers of the system to be localized without the use of any external tracking device. 

This work is organized as follows. The theory is presented in a general 3-D framework for matrix arrays in Section \ref{sec:theory}.  
The principles of plane wave (PW) imaging are summarized in Section \ref{sec:notation} along with the nomenclature used and the multiple transducer beamforming. Section \ref{sec:calculation-of-relative-position} describes the method for accurate calculation of the spatial location of the different transducers. Then the method is experimentally validated in 2-D using two identical linear arrays.
Experimental phantom measurements are described in Section \ref{sec:experiments}. The corresponding results, using the multi-transducer system, are shown Section \ref{sec:results}. 
To evaluate the potential gains in resolution and image contrast provided through our approach, all results are compared to the conventional PW imaging with one single transducer and the incoherently compounded image obtained with the multiple transducers. Finally, the implications of this work, including the limitations, are discussed in Section \ref{sec:discussion}. The study is concluded is Section \ref{sec:conclusion}.

\section{Theory}
\label{sec:theory}
Ultrasound image quality improves by reducing the F-number, which represents the ratio of the focusing depth to the aperture size. Expanding the aperture is a direct way to improve imaging performance. %Hence, if different transducers can be coherently combined, significantly increasing the aperture size of the system,  an enhanced image is expected. 
Preliminary in silico and phantom works suggest that different transducers can be coherently combined, significantly increasing the aperture size of the system and improving image resolution \cite{peralta2018feasibility,peralta2019SPIE}.

In the proposed coherent multi-transducer method, a single transducer is used for each transmission to produce a PW that insonates an entire FoV of the transmit transducer. The resulting echoes scattered from the medium are recorded using all the transducers that form the system. The sequence is continued by transmitting from each individual transducer in turn. Knowing the location of each transducer and taking into account the full transmit and receive path lengths, coherent summation of the radio frequency (RF) data from multiple transducers can be used to form a larger aperture and get an image, following the same approach as in PW imaging \cite{Montaldo2009CoherentElastography}.

\subsection{Multi-transducer notation and beamforming}
\label{sec:notation}
A 3-D framework consisting of $N$ identical matrix arrays (T$_i$, $i=1,\hdots,N$) that are freely placed in space and have a partly shared FoV is considered. The transducers are otherwise at arbitrary positions. 
All transducers are synchronized (i.e. trigger and sampling times in both transmit and receive mode are the same), and take turns to transmit a plane wave. Every transmitted wave is received by all transducers, including the transmit one. Thus, a single plane wave shot will yield $N$ RF datasets, one for each receiving transducer.

The framework is described using the following nomenclature. Points are noted in upper case letters (e.g. $P$), vectors representing relative positions are represented in bold lowercase (e.g. $\mathbf{r}$), unit vectors are noted with a ``hat'' (e.g. $\hat{x}$) and matrices are written in bold uppercase (e.g. $\mathbf{R}$). 
Index convention is to use $i$ for the transmitting transducer, $j$ for the receiving transducer, $h$ for transducer elements, and $k$ for scatterers. Other indices are described when used.

\begin{figure}[htb!]
\centering
\includegraphics[width=\linewidth]{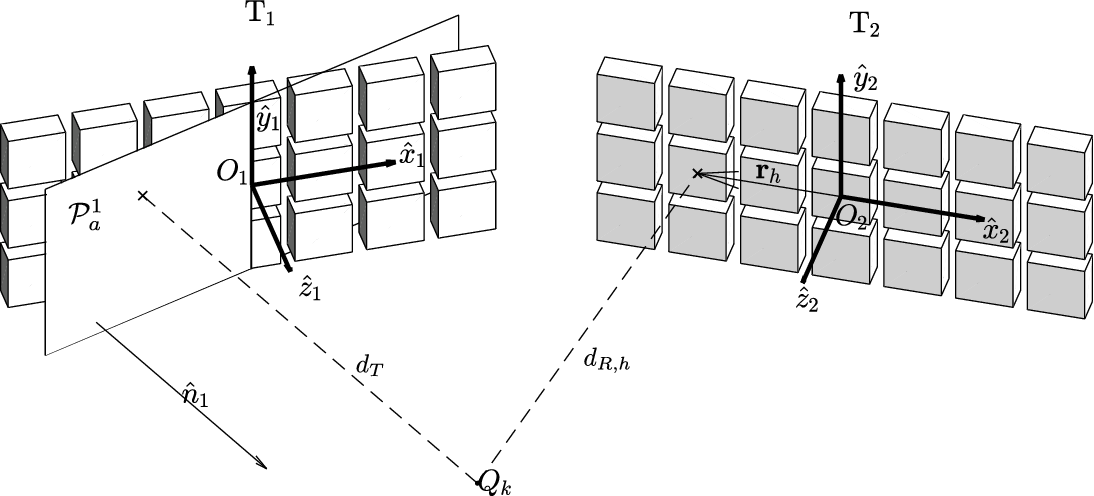}
\caption{Geometric representation of the multi-transducer beamforming scheme. In this example, transducer T$_1$ transmits a plane wave at certain angle defined by $\mathcal{P}_a^{1}$ and T$_2$ receives the echo scattered from $Q_k$ on element $h$.}
\label{fig:coordinate system}
\end{figure}

The set-up is illustrated in Fig. \ref{fig:coordinate system} for the simplest case of 2 transducers.
The resulting image and all transducer coordinates are defined in a world coordinate system arbitrarily located in space. The superscript $^i$ denotes when the transducer's local coordinates are used. 
The position and orientation of a transducer T$_i$ is represented by the origin $O_i$, defined at the center of the transducer surface, and the local axes $\{\hat{x}_i,\hat{y}_i,\hat{z}_i\}$, with the $\hat{z}_i$ direction orthogonal to the transducer surface and directed away from transducer $i$. A plane wave transmitted by transducer  T$_i$ is defined by the plane  $\mathcal{P}_a^{i}$, which can be characterized through the normal to the plane $\hat{n}_i$ and the origin $O_i$. The RF data received by transducer $j$ on element $h$ at time $t$ is noted $T_iR_j(h,t)$. 

Using the above notation, PW imaging beamforming \cite{Montaldo2009CoherentElastography} can be extended to the present multi-transducer set-up. Assuming that transducer T$_i$ transmits a plane wave at certain angle defined by $\mathcal{P}_a^{i}$, the image point to be beamformed located at $Q_k$ can be computed from the echoes received at transducer T$_j$ as:
\begin{equation}
\begin{split}
s_{i,j}(Q_k;\mathcal{P}_a^{i}) = \sum_{h=1}^{H} T_iR_j\left(h,t_{i,j,h}(Q_k;\mathcal{P}_a^{i})\right) = \\
\sum_{h=1}^{H} T_iR_j\left(h,\frac{D_{i,j,h}(Q_k;\mathcal{P}_a^{i})}{c}\right) 
\end{split}
\label{eq:beamforming}
\end{equation}
where $H$ is the total number of elements in the array, $c$ is the speed of sound of the medium, and $D$ is the distance travelled by the wave, which can be split into  the transmit and the receive distances:
\begin{equation}
D_{i,j,h}(Q_k;\mathcal{P}_a^{i}) = d_T(Q_k,\mathcal{P}_a^{i}) + d_{R,h}(Q_k, O_j+\mathbf{r}_h)
\end{equation}
with $d_T$ measuring the distance between a point and a plane (transmit distance), and $d_{R,h}$ being the distance between a point and the receive element (receive distance). These distances can be computed as follows:
\begin{equation}
d_T(Q_k,\mathcal{P}_a^{i}) =  | (Q_k-O_i) \cdot \hat{n}_i | =  | (Q_k - O_i) \cdot (\mathbf{R}_i \hat{n}_i^{i}) |
\end{equation}
and
\begin{equation}
\begin{split}
& d_{R,h}(Q_k,  O_j+\mathbf{r}_h) = \\
& \lVert Q_k - (O_j + \mathbf{r}_h) \rVert  = \lVert Q_k - ( O_j + \mathbf{R}_j \mathbf{r}^j_h )  \rVert 
\end{split}
\end{equation}
where $\lVert \rVert$ is the usual Euclidean distance,  and $\mathbf{R}_i = [\hat{x}_i \ \hat{y}_i \ \hat{z}_i]$ is a $3 \times 3$ matrix parameterized through three rotation angles, $\boldsymbol{\phi}_i = \{\phi_x, \phi_y, \phi_z\}_i$, that together with the offset $O_i$ characterize the position and orientation of transducer T$_i$ with 6 parameters \cite{fitzgibbon2003robust}.

With the total distances computed, equation (\ref{eq:beamforming}) can be evaluated for each pair of transmit-receive transducers, and the total beamformed image $S(Q_k)$ can be obtained by coherently adding the  individually beamformed images: 
\begin{equation}
S(Q_k;\mathcal{P}_a) = \sum_{i=1}^N \sum_{j=1}^N s_{i,j}(Q_k;\mathcal{P}_a^{i})
\label{eq:finalImage}
\end{equation}
In the same vein, assuming that the location of the multiple transducers of the system is known over the acquisition time and the medium of interest do not move, several plane waves transmitted at different angles, $a=1,\hdots,A$,  may be coherently combined as well to generate an image,
\begin{equation}
S(Q_k;\mathcal{P}_A) = \sum_{i=1}^N \sum_{j=1}^N \sum_{a=1}^A s_{i,j}(Q_k;\mathcal{P}_a^{i})
\label{eq:finalImageAngle}
\end{equation}

\subsection{Calculation of the transducer locations}
\label{sec:calculation-of-relative-position}

In order to carry out the coherent multi-transducer compounding described in the previous section, the position and orientation of each imaging transducers (defined by $O_i$ and $\boldsymbol{\phi}_i$) is required. This then allows computation of the travel time of the transmitted wave to any receive transducer.
This section describes a method to accurately calculate these positions by exploiting the consistency of received RF data when signals are received from the same medium insonated by different transducers.

A medium with $K$ point scatterers located at positions $Q_k$, ${k=1,\hdots,K}$ is considered. It is assumed that the speed of sound is constant and all transducers are identical (implications of these assumptions are discussed later in Section \ref{sec:discussion}).
The following transmit sequence is considered: a single plane wave is transmitted by each probe in an alternating sequence, i.e. only one probe transmits at each time while all probes receive, including the transmit one.  
Since the use of plane waves enables a high transmit rate, it can be assumed that the system remains still during consecutive acquisitions. Then, the wavefields resulting from the same point scatterer and received by the same transducer T$_j$, from consecutive transmissions by all transducers $T_{i=1,\cdots,N}$, must be correlated or have spatial covariance \cite{mallart1991van}. Specifically, considering only the RF data received by transducer T$_j$ (i.e. $T_{i=1,\cdots,N}R_j$), for each element $h$, any timing difference between them is the transmit time (receive time is equal since the receive transducer is the same). The signals received by element $h$ will be correlated once the difference in transmit time is compensated.
The proposed method consists of finding the optimal parameters for which the time correlation between the received RF datasets sharing the same receive transducer is maximum for $K$ scatterers in the common FoV. Those parameters define the total reception time corresponding to each point scatterer $Q_k$, and are:  
\begin{equation}
\mathbf{\theta} = \{\mathcal{P}_a^{1},...,\mathcal{P}_a^{N}, c, Q_1,\hdots,Q_K,\boldsymbol{\phi}_1,O_1,\hdots, \boldsymbol{\phi}_N,O_N\}
\end{equation}
\begin{figure*}[t!]
\scriptsize
% Store the current equation number.
\setcounter{MYtempeqncnt}{\value{equation}}
% Set the equation number to one less than the one
% desired for the first equation here.
% The value here will have to changed if equations
% are added or removed prior to the place these
% equations are referenced in the main text.
\setcounter{equation}{7}
\begin{equation}
%&\text{NCC}(E\{T_iR_j(h,t_{i,j,h}(Q_k;\mathcal{P}_a^{i})+T)\}, E\{T_jR_j(h,t_{j,j,h}(Q_k;\mathcal{P}_a^{j})+T)\}) = \nonumber \\
%&= \frac{\sum\limits_{\tau=0}^{T} (E\{T_iR_j(h,t_{i,j,h}(Q_k;\mathcal{P}_a^{i})+\tau)\}-\overline{E}\{T_iR_j(h,t_{i,j,h}(Q_k;\mathcal{P}_a^{i})+\tau))\}) (E\{T_jR_j(h,t_{j,j,h}(Q_k;\mathcal{P}_a^{j})+\tau)\}-\overline{E}\{T_jR_j(h,t_{j,j,h}(Q_k;\mathcal{P}_a^{j})+\tau)\})}{\left[\sum\limits_{\tau=0}^{T} (E\{T_iR_j(h,t_{i,j,h}(Q_k;\mathcal{P}_a^{i})+\tau)\}-\overline{E}\{T_iR_j(h,t_{i,j,h}(Q_k;\mathcal{P}_a^{i})+\tau)\})^2\sum\limits_{\tau=0}^{T}(E\{T_jR_j(h,t_{j,j,h}(Q_k;\mathcal{P}_a^{j})+\tau)\}-\overline{E}\{T_jR_j(h,t_{j,j,h}(Q_k;\mathcal{P}_a^{j})+\tau)\})^2\right]^{1/2}}
%
 \text{NCC}(E_{(i,j,h,k;a)}[T], E_{(j,j,h,k;a)}[T]) = 
 \frac{\sum\limits_{\tau=0}^{T} (E_{(i,j,h,k;a)}[\tau]-\overline{E}_{(i,j,h,k;a)}[\tau]) (E_{(j,j,h,k;a)}[\tau]-\overline{E}_{(j,j,h,k;a)}[\tau])}{\left[\sum\limits_{\tau=0}^{T}(E_{(i,j,h,k;a)}[\tau]-\overline{E}_{(i,j,h,k;a)}[\tau])^2\sum\limits_{\tau=0}^{T}( E_{(j,j,h,k;a)}[\tau]-\overline{E}_{(j,j,h,k;a)}[\tau])^2\right]^{1/2}} 
\normalsize
\label{eq:similarity1}
\end{equation}
% Restore the current equation number.
%\setcounter{equation}{\value{MYtempeqncnt}}
\hrulefill
% The spacer can be tweaked to stop underfull vboxes.
\vspace*{4pt}
\end{figure*}
\normalsize
Note that, in practice the angle of the transmitted plane wave is known and then the unknown parameters to optimize are the speed of sound and the locations of the scatterers and probes.
In addition, since the parameters that define transducer locations in space depend on the definition of the world coordinate system, the vector of unknown parameters can be reduced by defining the world coordinate system the same as the local coordinate system of one of the receiver transducers, e.g. T$_i$ $(\boldsymbol{\phi}_i=\{0,0,0\}, \; O_i=[0,0,0])$.

Being $T$ the time pulse length of the transmitted pulse, the envelope of the signal transmitted by  transducer T$_i$  backscattered by the scatterer $Q_k$ and received by transducer T$_j$, i.e. $T_iR_j(h,t_{i,j,h}(Q_k;\mathcal{P}_a^{i})+T)$  can be calculated as, 
\begin{equation}
\begin{split}
E_{(i,j,h,k;a)}[T] = 
E\{T_iR_j(h,t_{i,j,h}(Q_k;\mathcal{P}_a^{i})+T)\}= \\ 
\left[T_iR_j(h,t_{i,j,h}(Q_k;\mathcal{P}_a^{i})+T)^{2}+\right. \\
\left.\mathcal{H}\{T_iR_j(h,t_{i,j,h}(Q_k;\mathcal{P}_a^{i})+T)\}^{2}\right]^{1/2}
\end{split}
\end{equation}
where $\mathcal{H}$ is the Hilbert transform and to simplify the envelope of the signal is noted as $E_{(i,j,h,k;a)}[T]$.

Then, the similarity between signals received by the same element $h$ of transducer T$_j$ can be computed using equation (\ref{eq:similarity1}), where $\text{NCC}$ is the normalized crossed correlation.

Finally, the total similarity, $\chi_{j,k}$, between RF data received by the same transducer $j$ can be calculated taking into account all the elements as, 
%\begin{equation}
%\begin{split}
%& \chi_{j,k}(\mathbf{\theta}) =  \\
%& \sum_{i}^{N} \sum_{h}^{H} \text{NCC}(E\{T_iR_j(h,t_{i,j,h}(Q_k;\mathcal{P}_a^{i})+T)\}, \\
%& E\{T_jR_j(h,t_{j,j,h}(Q_k;\mathcal{P}_a^{j})+T)\}) W_{i,k,j,h} (\mathbf{\theta})W_{j,k,j,h}(\mathbf{\theta})
%\end{split}
%\end{equation}
\begin{equation}
\begin{split}
 \chi_{j,k}(\mathbf{\theta}) = &  
 \sum_{i}^{N} \sum_{h}^{H} \text{NCC}(E_{(i,j,h,k;a)}[T], \\
& E_{(j,j,h,k;a)}[T]) W_{i,k,j,h} (\mathbf{\theta})W_{j,k,j,h}(\mathbf{\theta}) 
\end{split}
\end{equation}
where $W_{i,k,j,h}$ is defined as,
\begin{equation}
\begin{split}
W_{i,k,j,h}(\mathbf{\theta}) &= \frac{1}{2} + \frac{1}{2H}\sum_{h_b \neq h}^{H} \text{NCC}(T_iR_j(h,t_{i,j,h}(Q_k;\mathcal{P}_a^{i})+T),  \\ 
& T_iR_j(h_b,t_{i,j,h}(Q_k;\mathcal{P}_a^{i})+T))  \\ 
& \textnormal{ with } h,h_b \in [1,\hdots,H]
\label{eq:weights}
\end{split}
\end{equation}  \normalsize

The function $W_{i,k,j,h}$ is a weighting factor proportional to the degree of coherence between pulses received across the individual elements of a single transducer, i.e. how well each signal correlates with those from the rest of the elements of the same transducer. 

% If the speed of sound is known, its value is 1 all along the same transducer array only when the ultrasonic wave is backscattered by a true point source in the medium, while the degree of similarity decreases as the estimated scatterer distance from the actual point source increases. In the same way, an under or overestimated speed of sound value will lead to a decrease of coherence.

Finally, summing over all receiving transducers of the system and scatterers yields the cost function:
\begin{equation}
\chi(\mathbf{\theta}) = \sum_j^{N} \sum_k^{K} \chi_{j,k}(\mathbf{\theta})
\end{equation}
Then, the optimal parameters $\bar{\theta}$, which include the relative position and orientation of all involved transducers, the speed of sound, and the position of the point scatterers can be found by a search algorithm that maximizes the cost function $\chi$,
\begin{equation}
\bar{\theta} = \arg \max_\theta \chi(\mathbf{\theta}) 
\label{eq:optimization}
\end{equation}
Equation (\ref{eq:optimization}) can be maximized by using gradient-based optimization methods \cite{lagarias1998convergence}.

Knowing the relative position of the different transducers of the system, the RF data can be beamformed using equation (\ref{eq:finalImage}). Note that, the world coordinate system where the multi-transducer image is reconstructed may be defined arbitrarily in space. A world coordinate system defined at the center of the total aperture of the system will lead to a more conventional point spread function (PSF), in which the best possible resolution is aligned with the lateral direction ($x$-axis) of the image. 

\subsection{Intuition and Uniqueness of solution}
In a homogeneous medium with $K$ point scatterers, the corresponding one-way geometric delay profile is a unique function of the target and array geometry and sound speed. Assuming a constant speed of sound in the medium, it is well known that the position of a point scatterer and the speed of sound can be estimated solely from the delays in the RF echo data recorded on individual elements of the receiver array \cite{anderson1998direct,desailly2013sono}. 

Given the position of a  number of point scatterers and with the assumption of a uniform speed of sound, the relative locations of the receivers and transmitters that form the imaging system can be calculated in similar way to trilateration positioning problems \cite{fang1986trilateration}. To localize a point, trilateration uses the location of at least three reference points (two points in 2-D) and the distance between them and the point to be localized.    
In 2-D geometry, it is known that if a point lies on two circles, then the circle centers and the two radii provide sufficient information to narrow the possible locations down to two. In 3-D geometry, when it is known that a point lies on the surfaces of three spheres, then the centers of the three spheres along with their radii provide sufficient information to narrow the possible locations down to no more than two. In both cases, 2-D and 3-D, additional information may narrow the possibilities down to one unique location.

In the context of the multi-probe system presented in this work. Once the relative position between a scatterer and the transducer T$_1$ are known, it is possible to estimate the distance between the scatterer and the second transducer T$_2$ through comparison of the RF data received by T$_1$, i.e., $T_1R_1$ and $T_2R_1$. Each additional point scatterer detected determines a sphere of center $Q_k = (x_k ,y_k ,z_k)$ and radius $d_T(Q_k;\mathcal{P}_a^{T_2})$. The location of transducer T$_2$ is determined relative to T$_1$ by one of the two external tangent planes common to three spheres defined by three different point scatterers. The direct RF echo data received by transducer T$_2$ i.e. $T_2R_2$ and $T_1R_2$ then provides the extra information required to determine the unique solution. In 2-D geometry two point scatterers provide the information required to solve this trilateration problem.

\section{Methods}
\label{sec:experiments}
The method was tested experimentally using 2 identical linear arrays having a partly shared FoV of an ultrasound phantom with both located on the same plane $(y=0)$. In this 2-D framework, the elevation dimension is removed from the problem and then the parameters that define the position and orientation of the transducers are reduced to one rotation angle $\{\phi\}$ and one 2-D translation $O_j$ \cite{fitzgibbon2003robust}. 
The experimental sequence starts with transducer 1 transmitting a plane wave into the region of interest (in the common FOV of transducer 1 and 2). Then, the backscattered ultrasound field  is received by both transducers of the system ($T_1R_1$ and $T_1R_2$). This sequence is repeated but transmitting with transducer 2 and acquiring the backscattered echoes with both transducers,  $T_2R_1$ and $T_2R_2$. Using this sequence, two different kind of experiments were carried out to validate the technique, a static configuration and a free-hand demonstration. 

\subsection{Experimental Setup}
The experimental setup was composed of two synchronized 256-channel Ultrasound Advanced Open Platform (ULA-OP 256) systems (MSD Lab, University of Florence, Italy) \cite{boni2016ula}. Each ULA-OP 256 system was used to drive an ultrasonic linear array made of 144 piezoelectric elements with a 6 dB bandwidth ranging from 2 MHz to 7.5 MHz (imaging transducer LA332, Esaote, Firenze, Italy). 
\begin{figure}[htb!]
\centering
\includegraphics[width=\linewidth]{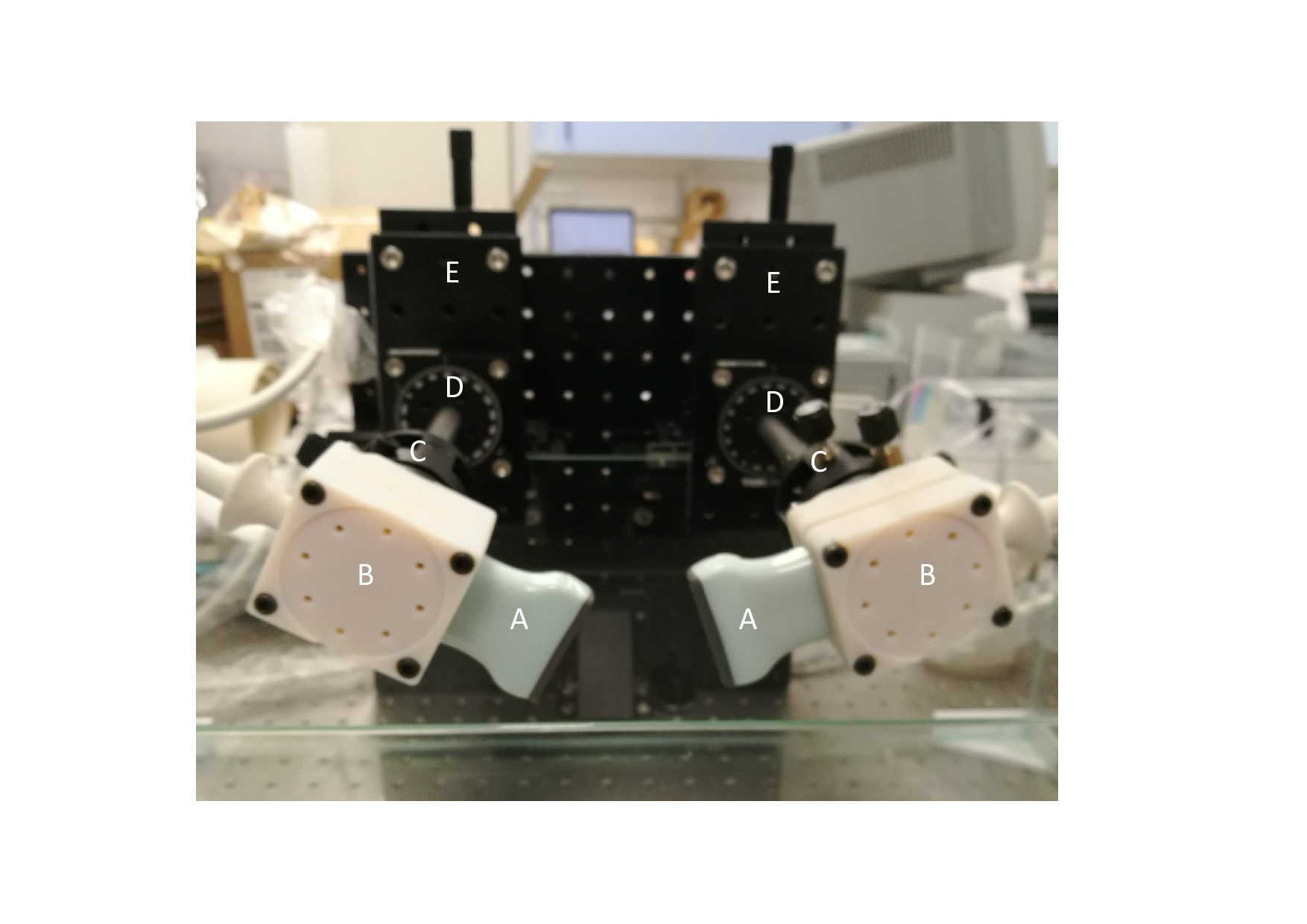} 
\caption{ Precision mechanical setup. Components are labeled with letters. (A) Linear array. (B) 3-D printed probe holder. (C) Double-tilt and rotation stage. (D) Rotation stage. (E)  $xyz$ translation stage.}
\label{fig:photosetup}
\end{figure}

Before acquisition, probes were carefully aligned in the same elevational plane using a  precision mechanical mount. Each probe was held by a 3-D printed shell structure that was connected to a double-tilt and rotation stage and then mounted on a $xyz$ translation and rotation stage (Thorlabs, USA). Fig. \ref{fig:photosetup} shows an annotated photograph of this setup. The imaging plane of both transducers $(y=0)$ was that defined by two parallel wires immersed in the water tank.

\subsection{Phantom}
\begin{figure}[htb!]
\centering
\includegraphics[width=\linewidth]{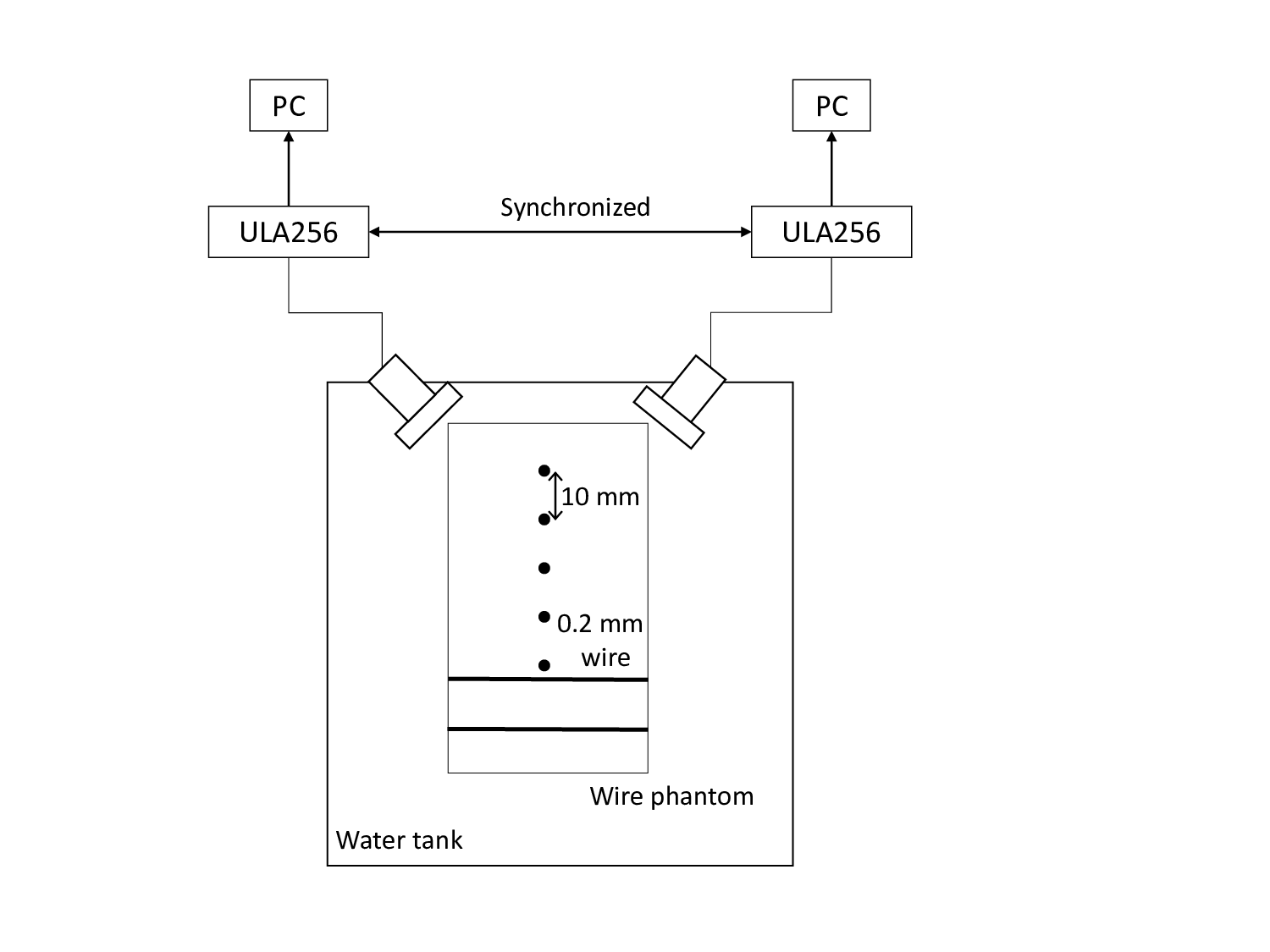} 
\caption{Experimental setup for coherent multi-transducer ultrasound imaging.}
\label{fig:setup}
\end{figure}

Two different ultrasound phantoms were used to experimentally validate the method and characterize resolution and contrast. 
The first phantom was a custom-made wire target phantom (200-$\mu m$ diameter) submersed in distilled water. Fig. \ref{fig:setup} shows a schematic view of this experimental setup.  

For measurement of contrast, an anechoic lesion phantom was produced. It was formed in a rectangular polypropylene mould of dimensions 13.5 cm x 10.2 cm x 18.5 cm. Two parallel walls of the mould (section 13.5 cm x 18.5 cm) were drilled to create a series of 3 holes in a line spaced $\sim$ 10 mm apart. Three nylon wires (200-$\mu m$ diameter) were passed through the holes and fixed. In line with these, a single cylindrical stainless steel bar (12.7 mm diameter, 102 mm) was also postioned (for later removal) to form the anechoic lesion. 

Then, 1200 mL deionized water was mixed with 28 g of agar until dissolved. The mixture was heated in a microwave oven up to 90$^o$ (boiling point of agar is 85$^o$).  When the solution reached the boiling point, it was removed from the oven and allowed to cool at room temperature while being stirred using a magnetic stirrer. When the temperature reached 50$^o$, 50 g of graphite were added without stopping stirring. The solution was then poured into the rectangular mould described above.

After room temperature was reached, the solution in mould was allowed to settle down in the fridge for at least 12h. Then the sample was carefully removed from the mould, keeping the wires and the bar embedded. In a final step, the stainless steel bar was removed from the sample. The resulting hole was filled with a similar agar mixture, except without graphite to make the anechoic lesion. 

During the experiments, the phantom was placed in a water tank at room temperature and positioned so that all wires and the anechoic region were in the common FoV of the 2 transducers. 

\subsection{Pulse sequencing and experimental protocol}
Two different kind of experiments were carried out. First, a stationary acquisition in which both probes were mounted and fixed in the precision mechanical setup described above. Resolution and contrast were measured in these conditions using the two ultrasound phantoms described above. The second experiment consisted of a free-hand demonstration. In this case, both probes were held and controlled by an operator. The transducer movements were carefully restricted to the same elevational plane, i.e. $y=0$ and to  keep two common targets in the shared FoV. To facilitate the alignment of the probes, the operator kept them in contact and parallel to the wall of the water tank. Data was acquired on the wire-phantom.

Three different types of pulse sequences were used. During the static experiment, to image the wire phantom and measure resolution, 121 plane waves, covering a total sector angle of 60$^{o}$  (from -30$^{o}$ to 30$^{o}$, 0.5$^{o}$ step),  were transmitted from the 144 elements of each probe at 3 MHz with a pulse repetition frequency (PRF)  equal to 4000 Hz. This is a total of 242 transmission events, since only one transducer transmits at each time while both probes simultaneously receive. The total sector angle between transmitted plane waves was chosen approximately the same as the angle defined between the probes. RF raw data backscattered up to 77 mm deep were acquired at a sampling frequency of 39 MHz. No apodization was applied either on  transmission or reception. The total time for this sequence was 60.5 ms.
Similar settings were used to measure contrast but transmitting only a single plane wave at 0$^{o}$ and at each time by both transducers.  The total time in this case for was 0.5 ms.

During the free-hand demonstration, 21 plane angles (from -5$^{o}$ to 5$^{o}$ with a 0.5$^{o}$  step) were transmitted from each probe and the backscattered signals from up to 55 mm deep were acquired. The rest of the settings were identical to the fixed probe experiment, i.e PRF of 4000 Hz and sampling frequency equal to 39 MHz. The total acquisition time using this sequence was 2 s, which results in 8000 transmit events in total.  Data was acquired only from the wire target phantom. 

\subsection{Data processing}
\label{sec_data_proc}
The initial estimate of the parameters, $\theta_{0}=\{c, Q_1,\hdots,Q_K,\phi_1,O_1, \phi_2,O_2\}$, needed to start the optimization algorithm was chosen as follows. 
Considering the world coordinate system the same as the local coordinate system of transducer T$_1$ ($\phi_1=0, \; O_1=[0,0]$), the parameter $\{\phi_2, O_2\}$ that define the position of transducer T$_2$ were calculated using point-based image registration \cite{beasley1999registration}. Two single images, T$_1$R$_1$ and T$_2$R$_2$, acquired by each of the transducers were used.  
For the scatterer positions $Q_k$ and  speed of sound of the propagation medium $c$, their initial value was calculated from the RF data T$_1$R$_1$ using the best-fit one-way geometric delay for the echoes returning from the targets, as described in \cite{anderson1998direct}.

Optimization was done using all the targets of the shared FoV. For the static experiment, since there is no motion, only one set of optimal parameters is needed and all RF data corresponding to plane waves transmitted at different angles can be beamformed using the same optimal parameters. However, to validate the optimization algorithm, 121 optimal parameter sets were calculated, one per transmit angle and using the same initial estimate.
On the other hand, for the free-hand demonstration, each frame was generated using a different set of optimal parameters, where after initializing the algorithm as described above, each subsequent optimization was initialized with the optimum value of the previous frame.

The proposed coherent multi-transducer method was compared with the image acquired using one single transducer and with the incoherent compounding of the envelope-detected images acquired by two independent transducers. The images acquired during the static experiment were used for this image performance analysis. A fully coherent image was obtained using equation (\ref{eq:finalImage}), by coherently adding the totality of the RF data acquired in one sequence (T$_1$R$_1$, T$_1$R$_2$, T$_2$R$_1$, T$_2$R$_2$):
\begin{equation}
\begin{split}
S(Q_k;\mathcal{P}_a) =  s_{1,1}(Q_k;\mathcal{P}_a^{1}) +  s_{1,2}(Q_k;\mathcal{P}_a^{1}) + \\
s_{2,1}(Q_k;\mathcal{P}_a^{2}) +  s_{2,2}(Q_k;\mathcal{P}_a^{2})
\end{split}
\end{equation}

Spatial resolution was calculated from the PSF on a single scatterer. An axial-lateral plane for 2-D PSF analysis was chosen by finding the location of the peak value in the elevation dimension from the envelope-detected data. Lateral and axial PSF profiles were taken from the center of the point target. The lateral and axial resolutions were then assessed by measuring the width and the axial (depth) of the PSF at the $-$6dB level, respectively. 
In addition, resolution was described using a frequency domain or k-space representation.  Axial-lateral RF PSFs were extracted from the beamformed data and the k-space representation was calculated using a 2-D Fourier transform. While the axial resolution is determined by the transmitted pulse length and the transmit aperture function, the lateral response of the system can be predicted by the convolution of the transmit and receive aperture functions \cite{walker1998application}. 

For the anechoic lesion phantom, the contrast and contrast-to-noise ratio (CNR) were measured from the envelope-detected images. Contrast was calculated as,
\begin{equation}
 \text{Contrast} = 20\log_{10}(\mu_i/\mu_o)
\end{equation}
 and CNR was computed as 
 \begin{equation}
 \text{CNR} = \mid \mu_i-\mu_o \mid / \sqrt{\sigma_{i}^{2}+\sigma_{o}^{2}}
\end{equation}
where $\mu_i$ and $\mu_o$ are the means of the signal inside and outside of the region, respectively, and $\sigma_{i}$ and $\sigma_{o}$ represent the standard deviation of the signal inside and outside of the region,
respectively.

\section{Results}
\label{sec:results}
\begin{figure*}[htb!]
\centering
\includegraphics[scale=1]{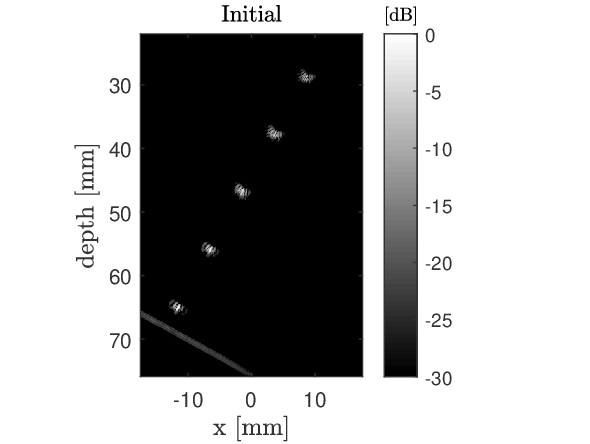}  \hspace{-3.75cm}
\includegraphics[scale=1]{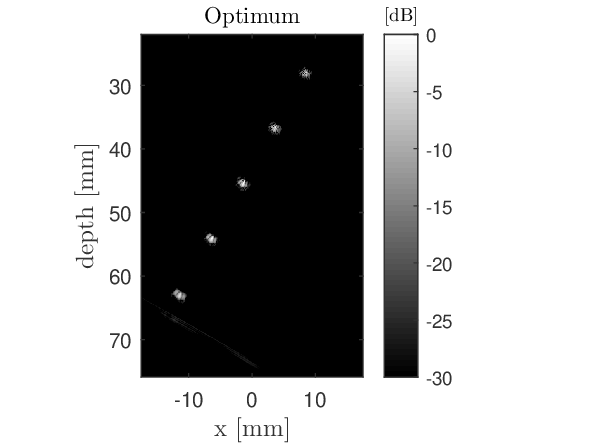} 
\caption{Experimental coherent multi-transducer images obtained using the initial estimate of the parameters ($\phi_2 = 55.33^{o}, O_2 = [39.55, 22.83]$ mm, $c=1496$ m/s) and their optimum values ($\phi_2 = 56.73^{o}, O_2 = [38.80, 23.06]$ mm, $c=1450.4$ m/s). Images formed compounding 121 angles over a total angle range of 60$^{o}$. Local coordinate system of transducer 1 used as world coordinate system for all images.}
\label{fig:optimization}
\end{figure*}

The 121 optimal parameter sets calculated for each of the transmit angles in the static experiment converged to the same solution. For the algorithm initialization, the estimated time from reconstructing the 2 images independently acquired by each probe, running the semi-automated registration method and optimizing the solution was less than 1 minute.
Fig. \ref{fig:optimization} shows the corresponding coherent multi-transducer images obtained using the initial estimate of the parameters and their optimum values. It is clearly shown that the blurring of the PSF presented in the image obtained using the initial estimate of the parameters is reduced after optimization. 

The convergence of the method was also validated in the free-hand experiment. In this case, each transmit angle was optimized over the total acquisition time. After calculating the initial estimate of the parameters of the first transmit PW as described in the previous section, each optimization was initialized with the optimum value of the previous transmission event.
As expected, rotation and translation parameters changed over acquisition time (following the operator movements), while the speed of sound can be considered constant. The averaged value and the standard deviation of the optimal speed of sound over the acquisition time was 1466.00 m/s $\pm$ 0.66 m/s. The resulting video, showing the sequence of succesfully optimized frames, can be found in the supporting material. \footnote{This paper has supplementary downloadable material available at http://ieeexplore.ieee.org, provided by the authors. This includes two multimedia AVI format movie clips, which show the results of the free-hand experiments.}

\begin{figure*}[htb!]
\centering
\includegraphics[width=\linewidth]{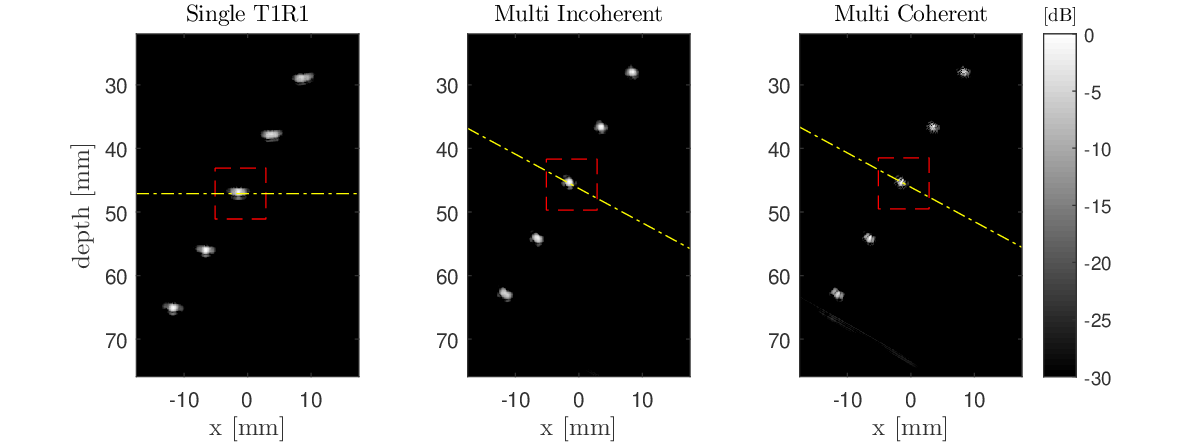} 
\caption{Experimental images of the wire phantom produced with a single transducer (T$_1$R$_1$), incoherently  compounding the images acquired by both transducers (envelope-detected images T$_1$R$_1$, T$_2$R$_2$) and coherently reconstructing the total RF data (using equation (\ref{eq:finalImageAngle})). Images formed compounding 121 angles over a total angle range of 60$^{o}$. Local coordinate system of transducer 1 used as world coordinate system for all images. PSF and transverse cut at the scatterer depth to estimate resolution are indicated with dashed lines. Note that, PSF and its cross section are calculated in the world coordinate system that leads to the best resolution in each case, i.e., conventional PSF of a single transducer calculated in the local coordinate system of transducer 1, and rotated by the bisector angle between transducers for the others PSFs.}
\label{fig:Bmode_T1R1vsT12R12_comp}
\end{figure*}

\begin{figure}[htb!]
\centering
\includegraphics[width=\linewidth]{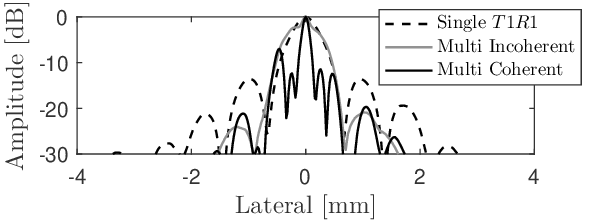} 
\caption{Transverse cut of the PSF at the scatterer depth defined in Fig. \ref{fig:Bmode_T1R1vsT12R12_comp} for the single transducer, incoherent and coherent multi-transducer methods. PSFs calculated in the world coordinate system that leads to the best resolution in each case, i.e., conventional PSF of a single transducer calculated in the local coordinate system of transducer 1, and rotated by the bisector angle between transducers for the others PSFs. For comparison, main lobes of the resulting transverse cuts are aligned within the lateral axis. Images formed transmitting a single plane wave at 0$^{o}$.}
\label{fig:resolution experiment}
\end{figure}

\begin{figure}[htb!]
\centering
\includegraphics[width=\linewidth]{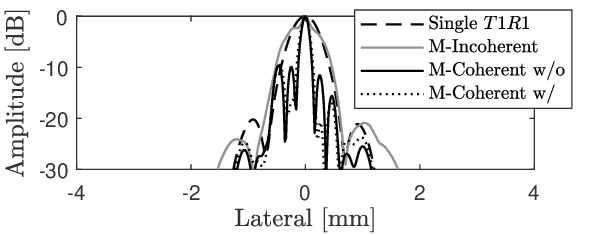} 
\caption{Transverse cut of the PSF at the scatterer depth defined in Fig. \ref{fig:Bmode_T1R1vsT12R12_comp} for the single transducer method and coherent multi-transducer method, with (w) and without (w/o) apodization. PSFs calculated in the world coordinate system that leads to the best resolution in each case, i.e., conventional PSF of a single transducer calculated in the local coordinate system of transducer 1, and rotated by the bisector angle between transducers for the others PSFs. For comparison, main lobes of the resulting transverse cuts are aligned within the lateral axis. Images formed compounding 121 plane waves over a total angle range of 60$^{o}$.}
\label{fig:resolution apodization}
\end{figure}

Fig. \ref{fig:Bmode_T1R1vsT12R12_comp} shows images for the wire phantom obtained using a single transducer (T$_1$R$_1$) , incoherently compounding the images acquired by both transducers (envelope-detected images T$_1$R$_1$, T$_2$R$_2$) and coherently reconstructing the total RF data (using equation (\ref{eq:finalImageAngle}))  after optimization. 
Comparing the resulting images between the case with a single transducer and the multi-transducer method, it is observed that the reconstructed images of the wire targets were clearly improved. 
The PSF of the three images were compared. Fig. \ref{fig:resolution experiment} and \ref{fig:resolution apodization} show the corresponding transverse cut of the PSF at the scatterer depth indicated by dotted lines in Fig. \ref{fig:Bmode_T1R1vsT12R12_comp} for each of these images, using a single PW at 0$^{o}$ and compounding 121 PW over a total angle range of 60$^{o}$, respectively. To analyze the multi-transducer method, the world coordinate system defined at the center of the total aperture, which leads to a more conventional PSF shape where the best resolution is aligned with the $x$-axis, is used. This coordinate system is defined rotating the local coordinate system of transducer T1 by the bisector angle between the two transducer, as indicated in Fig. \ref{fig:Bmode_T1R1vsT12R12_comp}. Also, note that, the incoherent multi-transducer results shown here benefit from the optimization, as the optimum parameters were used in the incoherent compounding of the enveloped-detected sub-images T$_1$R$_1$ and T$_2$R$_2$. 

The effect of the apodization on the multi-coherent PSF is presented in Fig. \ref{fig:resolution apodization}. The relative performance of all approaches is summarized in Table \ref{table:performance}.  The coherent multi-transducer acquisition presents the best lateral resolution, while the worst one corresponds to the incoherent image generated through combining the independent images acquired by both transducers. Also, larger differences are observed in the behavior of the side lobes, which are higher in the coherent multi-transducer method. When a single PW is used, the biggest difference is between the second side lobes, being raised by 13 dB for the coherent multi-transducer method compared to the single transducer method, while the difference of the first side lobes is 3.5 dB. This suggests that while significant image improvements can be achieved, the image may suffer from the effects of side lobes. The inclusion of the proposed apodization results in a significant reduction of the first side lobe and resolution improvement of 65\% compared to the conventional image acquired by a single transducer. In Fig. \ref{fig:Bmode_T1R1vsT12R12_comp} there is a noticeable variation in the PSF with increasing depth. This is due to the relative spatial position of the individual transducers and to the direction of the transmitted plane waves which determines the generation of the sidelobes in the reconstructed data.

\begin{table*}[ht!]
\caption{Imaging performance for the different methods.}
\label{table:performance} 
\centering 
\vspace{0.1cm}   
\begin{tabular}{lccccc}
\hline
 & Axial resolution [mm] &  Lateral resolution [mm]  & 1$^{st}$ sidelobe [dB] & 2$^{nd}$ sidelobe [dB] \\ \hline
 PW Conventional  (1 PW at 0$^{o}$) &0.9445 &0.6674  & -14.96& -20.79 \\
 Multi Incoherent (1 PW at 0$^{o}$) &0.9474 & 0.7837  & -20.87 & -\\
 Multi Coherent (1 PW at 0$^{o}$)  &0.8109 & 0.1817  &-11.46 & -7.01 \\
 \hline
%PW Conventional (121 PW, sector 60$^{o}$) &0.9002 &0.6546  & -20.22 & -\\
%Multi Coherent without apodization  (121 PW, sector 60$^{o}$) &0.8246 &0.1911  & -9.94 & -9.64\\
%Multi Coherent with apodization  (121 PW, sector 60$^{o}$)  &0.8391 &0.2278  & -20.73 & -9.45\\
 %\hline
\end{tabular}
\end{table*}

\begin{figure}[htb!]
\centering
\includegraphics[width=\linewidth]{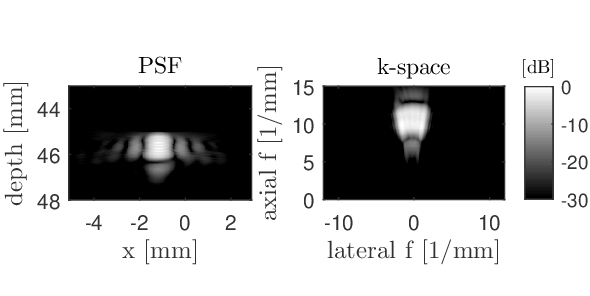} 
\includegraphics[width=\linewidth]{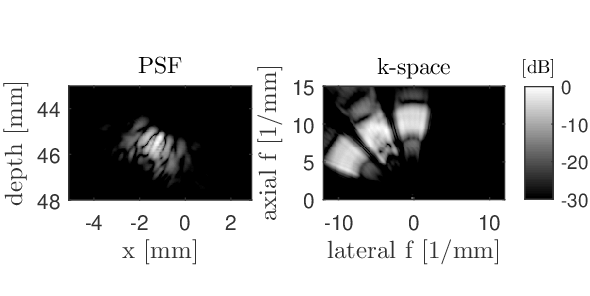} 
\caption{Envelope-detected PSFs and k-space representation obtained using a single transducer (upper graph) and using the coherent multi-transducer method (bottom graph). Images formed using a single PW at 0$^{o}$. Local coordinate system of transducer 1 used as world coordinate system.}
\label{fig:PSFKspace_angle0}
\end{figure}

\begin{figure}[htb!]
\centering
\includegraphics[width=\linewidth]{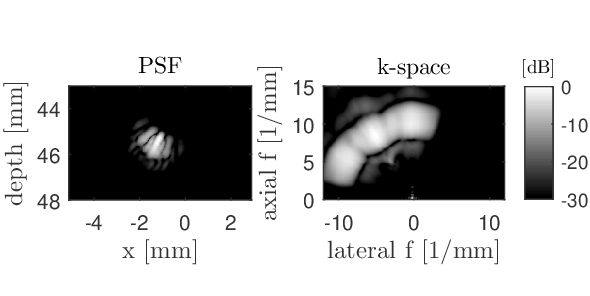} 
\includegraphics[width=\linewidth]{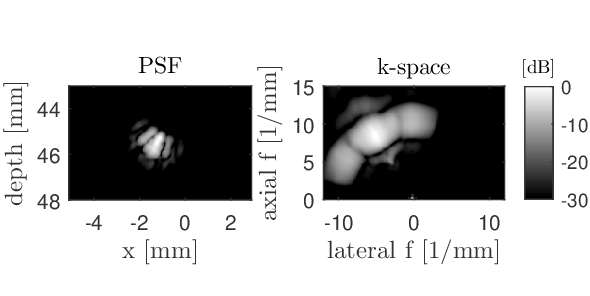} 
\caption{Envelope-detected PSFs and k-space representation of the multi-transducer method, compounding 121 plane waves covering a total angle range of 60$^{o}$, without (upper graph) and with apodization (bottom graph). Local coordinate system of transducer 1 used as world coordinate system.}
\label{fig:PSFKspace_coherent_local}
\end{figure}

The PSFs obtained using a single transducer and the coherently combined multi-transducer signals were investigated in k-space representation. Fig. \ref{fig:PSFKspace_angle0} shows the corresponding results using a single PW at 0$^{o}$. Images are represented in the local coordinate system of transducer 1. An important consequence of the linear system is that the superposition principle can be applied. As expected, the total k-space representation shows an extended lateral region that corresponds to the sum of the four individual k-spaces that form an image in the coherent multi-transducer method. It worth noting that, since both transducers are identical, they have the same k-space response (identical transmit and receive aperture functions) but in different k-space locations.  
The discontinuity in the aperture of the system, given by the separation between both transducers, leads to gaps in the spatial frequency space. This discontinuity can be filled by compounding PW over an angle range similar to the angle between by the two transducers. Since the lateral extent of k-space that can be reached with steered waves from a single transducer should be double that of the single plane wave (rectangle vs triangle function). Fig. \ref{fig:PSFKspace_coherent_local} shows the resulting PSF after compounding 121 angles with a separation of 0.5$^{o}$, which define a total sector of 60$^{o}$, and the corresponding continuous k-space. 
In addition, the topography of the continuous k-space can be re-shaped weighting the data from the different images that are combined to form the total one. A more conventional transfer function with reduced side lobes can be created accentuating the low lateral spatial frequencies, which are mostly defined by the sub-images T$_1$R$_2$ and T$_2$R$_1$. Using this approach, Fig. \ref{fig:PSFKspace_coherent_local} shows a PSF and its corresponding k-space representation generated weighting the sub-images T$_1$R$_1$, T$_1$R$_2$, T$_2$R$_1$ and T$_2$R$_2$ with the vector $[1, 2, 2, 1]$. Corresponding transverse cut of the PSF and imaging metrics are shown in Fig. \ref{fig:resolution apodization} and Table \ref{table:performance}.     

\begin{figure*}[htb!]
\centering
\includegraphics[scale=1]{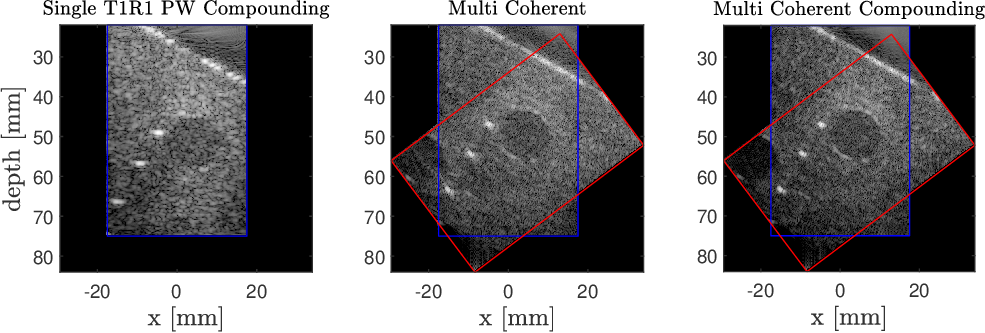}  
\caption{Experimental images of the contrast phantom produced coherently compounding 41 PW with a single transducer (T$_1$R$_1$), coherently compounding the RF data acquired by both transducers transmitting a single PW at 0$^{o}$ and transmitting 41 PW (using equation (\ref{eq:finalImageAngle})). The optimum parameters used to reconstruct the multi-coherent image are $\phi_2 = 53.05^{o}, O_2 = [41.10, 25.00]$ mm, $c=1437.3$ m/s.}
\label{fig:contrast}
\end{figure*}

The results obtained from the anechoic lesion phantom are presented in Fig. \ref{fig:contrast}.  The initial estimate of the parameters was chosen as described in Section \ref{sec_data_proc} and the 3 strong scatterers generated by the nylon wires were used in the optimisation. It can be seen that, in general, the multi coherent image has better defined edges, making a border easier to delineate than the image obtained by a single transducer. The reconstructed images of the wire targets were clearly improved, the speckle size was reduced and the anechoic region was easily identifiable from the phantom background. The lesion was visible with a contrast of -6.708 dB and a CNR of 0.702 in the single transducer image, while those values for the multi coherent image were -7.251 dB and 0.721, respectively.
%\begin{figure*}[htb!]
%\centering
%\includegraphics[scale=1]{figures/peral10a}  \hspace{-3.75cm}
%\includegraphics[scale=1]{figures/peral10b} 
%\caption{Experimental images of the contrast phantom produced with a single transducer (T$_1$R$_1$) and coherently compounding the images acquired by both transducers  (using equation (\ref{eq:finalImage})). The optimum parameters used to reconstruct the multi-coherent image are $\phi_2 = 53.05^{o}, O_2 = [41.10, 25.00]$ mm, $c=1437.3$ m/s.}
%\label{fig:contrast}
%\end{figure*}
%\begin{figure*}[htb!]
%\centering
%\includegraphics[scale=1]{new_figures/Images_41PW_vs0PW}  
%\caption{\hl{Experimental images of the contrast phantom produced coherently compounding 41 PW with a single transducer (T$_1$R$_1$), coherently compounding the RF data acquired by both transducers transmitting a single PW at 0$^{o}$ and transmitting 41 PW (using equation} (\ref{eq:finalImageAngle})). \hl{The optimum parameters used to reconstruct the multi-coherent image are $\phi_2 = 53.05^{o}, O_2 = [41.10, 25.00]$ mm, $c=1437.3$ m/s.}}
%\label{fig:contrast}
%\end{figure*}
%\begin{figure*}[htb!]
%\centering
%\includegraphics[scale=1]{new_figures/Fov_T1vsCMTUS_41PW}  
%\caption{\hl{Experimental images of the contrast phantom produced coherently compounding 41 PW with a single transducer (T$_1$R$_1$), coherently compounding the RF data acquired by both transducers transmitting a single PW at 0$^{o}$ and transmitting 41 PW (using equation} (\ref{eq:finalImageAngle})). \hl{The optimum parameters used to reconstruct the multi-coherent image are $\phi_2 = 53.05^{o}, O_2 = [41.10, 25.00]$ mm, $c=1437.3$ m/s.}}
%\label{fig:contrast}
%\end{figure*}

\begin{table*}[ht!]
\caption{Imaging performance for the different methods assessed on the contrast phantom.}
\label{table:inclusionPerformance} 
\centering 
\vspace{0.1cm}   
\begin{tabular}{lcccc}
\hline
 & Lateral resolution [mm] & Contrast [dB] & CNR [-] & Frame rate [Hz] \\ \hline
 Single T1R1 (1 PW at 0$^{o}$) & 2.633 & -6.708 & 0.702 & 10700 \\
Single T1R1 Compounding (41 PW, sector 20$^{o}$) & 1.555 & -8.260 & 0.795 &  260 \\
 \hline
 Multi Coherent (1 PW at 0$^{o}$) & 0.713 & -7.251 & 0.721 & 5350 \\
Multi Coherent Compounding (41 PW sector 20$^{o}$)  & 0.693 & -8.608 & 0.793 & 130 \\
\hline
\end{tabular}
\end{table*}

\begin{figure}[htb!]
\centering
\includegraphics[width=\linewidth]{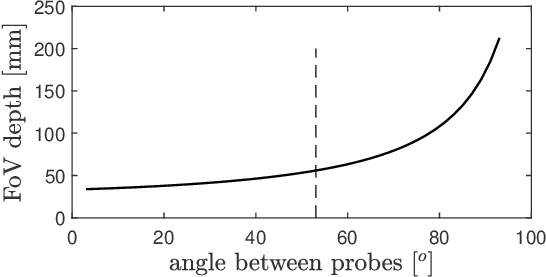}  
\caption{Depth of common field of view as function of the angle between both probes when transmitting plane waves at 0$^{o}$ and keeping the rest of the parameters that define the position of the probes in space constant and equal to the one used to acquire the images of Fig.} \ref{fig:contrast}, i.e. $O_2 = [41.10, 25.00]$ mm.
\label{fig:fov}
\end{figure}

\section{Discussion}
\label{sec:discussion}
This study introduces a new coherent multi-transducer ultrasound system that significantly outperforms single transducer through coherent combination of signals acquired by different synchronized transducers that have a shared FoV. Since the multi coherent image is formed by 4 RF datasets that are acquired in two consecutive transmissions, it is necessary that tissue and/or probe motion do not break the coherence between consecutive acquisitions. To ensure this is the case high frame rate acquisition is essential. Its performance has been demonstrated using plane waves. However, different transmit beam profiles such as diverging waves may increase the overlapped FoV, extending the final high resolution image. 
The issue of FoV and resolution gain as probes are moved apart is very real. In our method there needs to be sufficient overlap of insonated regions to allow the relative probe positions to be determined. Any overlapping regions will benefit from improved resolution as a result of the enlarged aperture of the combination of transducers. Other regions that extend the FoV with no overlap clearly will simply achieve an extended FoV without a resolution gain. Thus we see this scenario as providing net benefits, but of different kinds in different locations.
The experiments presented here were performed as a demonstration in 2-D, using linear arrays. To facilitate this, the linear array transducers were precisely aligned in the elevation plane. The framework that we propose clearly encompasses the 3rd spatial dimension. In future work the use of matrix arrays capable of volumetric acquisitions could be used for a true 3-D demonstration.

In relation to current ultrasound imaging systems, the use of multiple probes will potentially increase the operational difficulty for the individual performing the scan. However, it is possible to manipulate multiple probes using a single, potentially adjustable, multi-probe holder that would allow the operator to hold multiple probes with only one hand while keeping directed to the same region of interest. In related work, such a probe holder has been demonstrated as a potential device for incoherent combination of multiple images for extended FoV imaging  \cite{zimmer2018}. The approach presented in this study could lead to a totally different strategy in US in which large assemblies of individual arrays are operated coherently together.
%\hl{On the other hand, while additional probes increase the complexity and cost, ultrasound systems with high channel numbers are becoming more widely available} \cite{hager2016ekho, provost20143d} \hl{and in the future it is likely that transducer technologies such as the proposed coherent multi-transducer ultrasound system might lead to more affordable and versatile probe designs} \cite{tanter2014ultrafast, kurth2016mobile}.

To successfully improve the PSF, the proposed multi-transducer method requires coherent alignment of the backscattered echoes from multiple transmit and receive positions. This requirement is only achieved through precise knowledge of all the transducer positions, which in practice is not achievable by manual measurements or using electromagnetic or optical trackers \cite{mercier2005review}. This study, presents a method for precise and robust transducer location based on the coherence of backscattered echoes arising from the same point scatterer and received by the same transducer using sequential transmissions from each of the transducers of the system. The location of the transducers is calculated by optimizing the coherence between individual receive elements calculated by cross-correlating the backscattered of common target signals.
Like in free-hand tracked ultrasound for image guide applications \cite{najafi2014closed, boctor2004novel}, spatial calibration is essential to guarantee the performance of the proposed multi-coherent ultrasound method.
The use of gradient-descent methods requires an initial estimate of the parameters close enough to the global maximum of the cost function. The distance between maxima, which depends on the NCC and corresponds to the pulse length, dictates this tolerance. This is approximately 1.5 $\mu$s (equivalent to 2.19 mm) for the experimental configuration used here. This tolerance value can be realistically achieved through image registration \cite{beasley1999registration}. In practice, in a free-hand situation, and assuming that at some initial instant the registration is accurate, this initial guess can be ensured if the transducers move relatively little in the time between two transmissions and share a common FoV. In PW imaging, the frame rate is only limited by the round-trip travel time, which depends on the speed of sound and the depth. For the experimental setup used in this work, the minimum time between two insonifications is around 94 $\mu$s. Hence the maximum frame rate is limited to $F_{max} = 10.7$ kHz, which in the case of the present multi transducer coherent method is reduced by the number of probes as $F_{max}/N$. 
To guarantee free-hand performance of the multi transducer method, perfect coherent summation must be achieved over consecutive transmissions of the $N$ transducers of the system. However, when the object under insonification moves between transmit events, this condition is no longer achieved. In other words, the free-hand performance is limited by the maximum velocity at which the probes move. Considering that coherence breaks for a velocity at which the observed displacement is larger than half a pulse wavelength per frame \cite{denarie2013coherent}, the maximum velocity of the probes is $V_{max} = \lambda F_{max}/2N$, which in the example showed here is 1.33 m/s.  This speed far exceeds the typical operator hand movements in a regular scanning session and hence, the coherent summation over two consecutive transmission is achieved. 
The method has been validated in a free-hand demonstration. 

Nowadays, in clinical practice the aperture is limited because extending it often implies increasing the cost and the system complexity. This work uses conventional equipment and image-based calibration to extend the effective aperture size while increasing the received amount of RF data (data x $N$). The estimated time for the first initialization is less than 1 minute, which is comparable to other calibration methods \cite{najafi2014closed, boctor2004novel}. Once the algorithm has been correctly initialized, the subsequent running times for the optimization can be significantly decrease. For example, in the free-hand experiment, where each optimization was initialized with the output from the previous acquisition. Finally, similar to 3-D and 4-D ultrafast imaging where the data is significantly large \cite{provost20143d}, in the proposed multi-transducer method computation may be a bottleneck for real time imaging. Graphical processing unit (GPU)-based platforms and high-speed buses are key to future implementation of these new imaging modes \cite{tanter2014ultrafast}.

Our results suggest that the improvements in resolution are mainly determined by the achieved extended aperture rather than compounding PW at different angles. In the coherent multi-transducer method there is a trade-off of between resolution and contrast. While a large gap between the probes will result in an extended aperture and improvement in resolution, contrast may be compromised due to the effects of sidelobes. The differences between the k-space representations for the single and the coherent multi-transducer methods further explain the differences in imaging performance; the more extended the k-space representation, the higher the resolution \cite{anderson2000seminar}. The relative amplitudes of the spatial frequencies present, i.e. the topography of k-space, determine the texture of imaged targets. Weighting the individual data from the different transducers can reshape the k-space, accentuating certain spatial frequencies and so can potentially create a more conventional response for the system. Moreover, the presence of uniformly spaced unfilled areas in a system's k-space response may indicate the presence of grating lobes in the system's spatial impulse response \cite{walker1998application}. A sparse array (like our multi-transducer method) creates gaps in the k-space response. Only with minimal separation between transducers the k-space magnitude response will become smooth and continuous over an extended region. This suggests that there is an interplay between the relative spatial positioning of the individual transducers and the angles of the transmitted planes waves; where either one or both of these can determine the resolution and contrast achievable in the final image. There is an opportunity to use the relative position data to decide what range of PW angles to use and to change these on the fly to adaptively change performance. Accordingly, in the free-hand experiment, the maximum transmit angle was limited to maintain a common FoV over time while the probes are moving. Finally, in real life applications, resolution and contrast will be influenced by a complex combination of probe separation and angle, aperture width, fired PW angle and imaging depth. In the future, we will focus on further investigating these different factors that determine the image performance of the system \cite{peralta2019SPIE}.

Image enhancements related to increasing aperture size are well described \cite{bottenus2017evaluation}. Nevertheless, large-aperture arrays represents ergonomic operator problems and have limited flexibility to adapt to different applications.  
In this work, the extended aperture is the result of adding multiple freely placed transducers together, which allows more flexibility. Small arrays are easy to couple to the skin and adapt to the body shape. In related work \cite{bottenus2016feasibility}, an effective extended aperture was created by combining synthetic aperture data sets over a range of aperture positions  while precisely tracking the position and orientation of the transducer \cite{bottenus2016feasibility}. But unlike the method presented here, success relies on the accuracy of the tracking system and ultrasound calibrations \cite{zhang2016synthetic}. Furthermore, the scan time to form a coherent aperture compounding over different aperture positions is high and in practice limited by tissue motion \cite{bottenus2017evaluation}. A key feature to consider in the coherent multi-transducer system presented in this paper is that all probes receive simultaneously reducing effect of tissue motion during reception.  In addition, since all elements are used in transmission, the use of plane wave generates a higher energy wavefield than in the synthetic aperture approach, improving penetration, and also enabling higher frame rates \cite{Montaldo2009CoherentElastography}.    

Wavefront aberration caused by inhomogeneous medium can significantly limit the quality of medical ultrasound images and is the major barrier to achieve diffraction-limited resolution with large aperture transducers \cite{lacefield2002distributed}. The technique described in this work has been tested in a scattering medium, with the assumption of a constant speed of sound along the propagation path. However, since the speed of sound is a parameter in the optimization, the technique could be adapted for nonhomogeneous media where the speed of sound varies in space. In this case, the medium could be modeled through piecewise continuous layers. The optimization method could be applied in a recursive way, dividing the FoV in sub areas with different speeds of sound. More accurate speed of sound estimation would improve beamforming and allow higher order phase aberration correction. Potentially representing speed of sound maps would be of great interest in tissue characterization \cite{bamber1981acoustic, imbault2017robust}. In addition, the use of multiple transducers allows multiple interrogations from different angles, which might give insight into the aberration problem and help to test new algorithms to remove the clutter.  

Further studies are needed to predict the performance of the proposed multi transducer system for in-vivo imaging. The approach presented here has been formulated and validated for detectable and isolated point scatterers within the shared imaging region, which in practice may not be always possible. However, although the theory was presented for point-like scatterers, the approach relies on a measure of coherence which may well be more tolerant, as indicated in the contrast phantom demonstrated in Fig. \ref{fig:contrast}. This result suggests that the method may work when there are identifiable prominent local features, and the concept of maximizing coherence of data received by each receiver array when insonated by different transmitters could allow wider usage. Indeed, an optimization based on spatial coherence might be more robust in the case where point targets are not available, due to the expected decorrelation of speckle with receiver location\cite{bottenus2015acoustic,liu1995application,walker1997speckle}. This may also lead to improvements in computational efficiency. Measures of spatial coherence have been used previously in applications such as phase aberration correction \cite{liu1998estimation}, flow measurements \cite{li2015coherent}, and beamforming \cite{lediju2011short}. %\hl{Exploring a new approach mainly based on spatial coherence will be the focus of future works on coherent multi-transducer imaging.}
On the other hand, isolated point scatterers can be artificially generated by other techniques, for instance by inclusion of microbubble contrast agents. Recently, ultrasound super-resolution imaging has demonstrated that spatially isolated individual bubbles can be considered as point scatterers in the acoustic field \cite{christensen2015vivo} and accurately localized \cite{christensen2017microbubble}. The feasibility of the coherent multi-transducer method in complex media, including a new approach mainly based on spatial coherence \cite{mallart1991van,liu1995application} and the potential use of microbubbles \cite{peralta2019ICUS} will be the focus of future work.

\section{Conclusion}
\label{sec:conclusion}

In this study a new coherent multi-transducer ultrasound imaging system and a robust method to accurately localize the multiple transducers have been presented. The subwavelength localization accuracy required to merge information from multiple probes is achieved by optimizing the coherence function of the backscattered echoes coming from the same point scatterer insonated by sequentially all transducers and received by the same one, without the use of an external tracking device. The theory for the approach was general for a multiplicity of 2-D arrays placed in 3-D and the method was experimentally validated in a 2-D framework using a pair of linear array and ultrasound phantoms. The improvements in imaging quality have been shown. Overall the performance of the multi-transducer approach is better than PW imaging with one single linear array. Results suggest that the coherent multi-transducer imaging has the potential to improve ultrasound image quality in a wide range of scenarios.

\section*{Acknowledgment}
This work was supported by the Wellcome/EPSRC Centre for Medical Engineering [WT 203148/Z/16/Z] and the Wellcome Trust IEH Award [102431]. The authors acknowledge financial support from the Department of Health via the National Institute for Health Research (NIHR) comprehensive Biomedical Research Centre award to Guy's \& St Thomas' NHS Foundation Trust in partnership with King's College London and King’s College Hospital NHS Foundation Trust. The views expressed are those of the authors and not necessarily those of the NHS, the NIHR or the Department of Health.

\ifCLASSOPTIONcaptionsoff
  \newpage
\fi



\bibliographystyle{IEEEtran}
\end{document}